\documentclass[12pt]{iopart}

\usepackage{graphicx}
\usepackage{cite}
\usepackage[colorlinks, citecolor = blue, urlcolor = blue]{hyperref}
\def\mydoubleq#1{``#1''}

\begin{document}

\title[Quenched Freely Jointed Chain]{Elasticity of a Freely Jointed Chain with Quenched Disorder}

\author{Minsu Yi and Panayotis Benetatos}
\address{Department of Physics, Kyungpook National University \\ 80 Daehak-ro, Buk-gu, Daegu 41566, Korea}
\ead{alstn8588@knu.ac.kr, pben@knu.ac.kr}
\vspace{10pt}
\begin{indented}
\item[]March 2025
\end{indented}

\begin{abstract}
We introduce a simple theoretical model, the Freely Jointed Chain with quenched hinges (qFJC), which captures the quenched disorder in the local bending stiffness of the polymer. In this article, we analyze the tensile elasticity of the qFJC in the Gibbs (fixed-force) ensemble. For finite-size systems, we obtain a recurrence relation of the exact free energy, which allows us to calculate the exact force-extension relation numerically for an arbitrary size of the system. In the thermodynamic limit, when $L({\rm contour \;length})\gg L_p({\rm persistence \;length})$, we obtain a framework to deal with quenched disorder in the polymer configuration. This allows us to obtain the response function for the discrete and continuous qFJC in the thermodynamic limit. It turns out that the extension of the continuous qFJC can be cast in a simple form. Furthermore, we have applied our analysis to rod-coil multiblock copolymers.
\end{abstract}

%
% Uncomment for keywords
%\vspace{2pc}
%\noindent{\it Keywords}: XXXXXX, YYYYYYYY, ZZZZZZZZZ
%
% Uncomment for Submitted to journal title message
%\submitto{\JPA}
%
% Uncomment if a separate title page is required
%\maketitle
% 
% For two-column output uncomment the next line and choose [10pt] rather than [12pt] in the \documentclass declaration
%\ioptwocol
%

\section{\label{sec:intro}Introduction}

Quite often, the macroscopic behavior of a system is affected by disorder, and the type of disorder plays an important role. Thus, the difference between the types of disorder associated with certain degrees of freedom has been of interest to various systems, including macromolecules, both synthetic and biopolymers. The most well-known and also well-studied problem is that of polypeptides that undergo helix-coil transition. In the helix state, the monomers form a rotating pattern induced by hydrogen bonds that increase the bending stiffness, whereas, in the coil state, the monomers are oriented randomly, which can be viewed - for zero or small tension - as a Gaussian chain.

In~\cite{Buhot_PRL}, the mean-field version of the free energy density and the force-extension relation were analyzed, for annealed disorder, treating the helix state as a longer rod (infinite bending stiffness) and the coil state as a freely jointed chain. The quenched-disorder case was also considered, where the freely jointed chain consists of monomers (rods) with two different lengths. The difference in the monomer length is the quenched disorder.~\cite{Tamashiro_Helix-coil_PhysRevE} treats the annealed version of the same model exactly and points out that there are stress-induced helix-coil crossovers for finite values of the cooperativity parameter, where the transition to the helix state stiffens the chain. A genuine phase transition appears only in the fully cooperative case.~\cite{Debnath_AB_block} considered semi-flexible A-B block copolymers under tension, also treating both types of disorder. The semi-flexible blocks are modeled as wormlike-chain (WLC) segments with different persistence lengths. The approach was based on the Markov process method introduced by Fredrickson, Milner, and Leibler in~\cite{Fredrickson:1992aa}. The same approach, applied to randomly flexible heteropolymers without tension, was presented in~\cite{Debnath_free}. A numerical study of the conformational and elastic properties of the Kratky-Porod chain with quenched disorder in the bending stiffness was presented in~\cite{Muhuri_2010}.

Apart from studies of polymers with random local bending stiffness, other types of quenched disorder along the polymer contour have also been explored. For example, since the base pairs on the dsDNA differ along the contour, the corresponding curvature can be viewed as a random variable, implying that the architecture is quenched. Using the framework of the WLC and the replica trick~\cite{Edwards:1975aa}, the force-extension relation and the transverse fluctuations can be analytically computed for such biopolymers~\cite{PB2010QuenchedWLC}. The effect of quenched random transverse forces on the tensile elasticity of a WLC was analyzed in~\cite{PB2010QuenchedWLC,PB_random_force_constrained}. In addition, methods from the statistical physics of disordered systems ({\it e.g.}, spin glasses) have been applied to protein folding and random heteropolymers (random hydrophobic-hydrophilic chain, random bond cain, random sequence chain)~\cite{Orland_in_Young}.

In a recent work, we have analyzed the stretching and bending elasticity of the freely jointed chain with reversible hinges (rFJC) in the Gibbs ensemble~\cite{Noh_rFJC, Yi_rFJC}, which is the annealed scenario of the model we will present in this article. Actually, this model (the annealed version) was originally introduced and analyzed in the Helmholtz (fixed extension) ensemble by Winkler {\it et al.}~\cite{Winkler_FJC, Winkler_FJC_macromolecules}

This model is the simplest model that captures the effect of a random (annealed) bending stiffness. An open hinge acts the same as in the usual freely jointed chain (uFJC), and a closed hinge links the two adjacent segments (which are rigid rods), forming a longer rod with infinite bending stiffness. In the thermodynamic limit of the rFJC, it turns out that the force-extension relation in the large force limit becomes the form of \(1-{\rm e}^{-f}\), {\it i.e.} exponential decay in the differential tensile compliance. This is qualitatively different from the uFJC, where the compliance decays as a power law.

The simplest way to introduce quenched disorder in the local bending stiffness is to make an analogy with the rFJC. This can be accomplished by letting the reversible hinges of the rFJC get quenched. This is the main topic of the paper, which is the freely jointed chain with quenched hinges. Then, one can ask the following questions: How will the force-extension relation differ from the uFJC or the rFJC? Is there a general framework to deal with quenched disorder on polymers? In this article,  we try to gain insight into these questions.

The article is organized as follows. In Section~\ref{sec:model}, we introduce the model of a freely jointed chain with quenched hinges in detail. In Section~\ref{sec:exact}, we discuss a simple case of the model with one random hinge and obtain a recurrence relation for the exact mean free energy. In Section~\ref{sec:thm_lim}, we derive a mathematical form to deal with the thermodynamic limit for quenched-disorder polymers and obtain the force-extension relation for the discrete and continuous models. Finally, in Section~\ref{sec:general}, we obtain a force-extension law for random copolymers and show an example of how this approach can be implemented.

\section{\label{sec:model}Model}

We start by considering a freely jointed chain consisting of \(N+1\) chain segments, each of which is a one-dimensional massless rigid rod of length \(b\), where the segments are connected by hinges \(N\). The total contour length is  \(L=(N+1)b\). We consider the chain to be in three dimensions, allowing for any orientation of the segments along the unit sphere.

To introduce quenched disorder on the hinges, we assign a uniform closing (occupation) probability \(p\) to each hinge. If the hinge is open (unoccupied), the two connected segments can have any relative orientation with uniform probability. When a hinge is closed (occupied), the two connected segments align and act as a longer rigid rod. Note that fluctuations on the hinge state are not allowed, {\it i.e.}, the hinge state is frozen over time. Since the disorder of the hinges is quenched, this captures the non-fluctuating (quenched) local bending stiffness.

\begin{figure}
    \centering
    \includegraphics[width=0.8\linewidth]{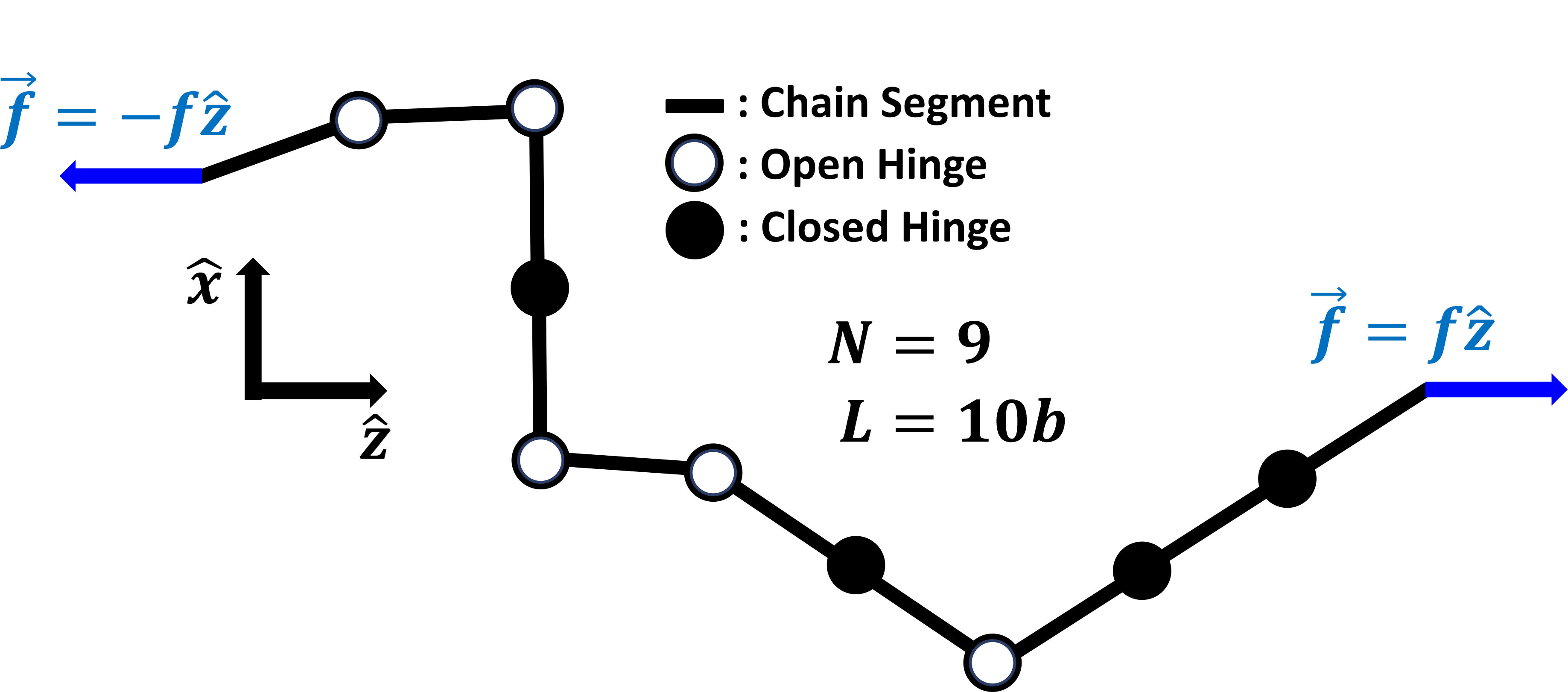}
    \caption{One realization of the qFJC under tensile force.}
    \label{fig:qFJC_intro}
\end{figure}

In this paper, we analyze the behavior of this chain in the Gibbs ensemble (fixed-force ensemble, where the control parameter is the applied force, and the extension can fluctuate), where a constant stretching force is applied at the two ends of the chain. An example of one realization of the model is shown in figure~\ref{fig:qFJC_intro}. Before we go further, for the sake of simplicity, we define a new abbreviation, the qFJC, which stands for the Freely Jointed Chain with quenched disorder on the state of the hinges.

A microstate of the chain can be expressed in terms of the orientation with respect to the direction \(\hat{z}\), \(\Omega_i=\{\theta_i,\phi_i\}\) for each chain segment (\(i=1,2,\dots, N+1\)). We also define \(n_i=0\) or \(1\) as the occupation number, which serves to specify the open or closed state of the \(i\,\)th hinge. If we choose \(\hat{z}\) as the direcction of the applied force, so that \(\vec{f}=f\hat{z}\), the Hamiltonian of the system can be written as
\begin{eqnarray}
    H = -\sum_{i=1}^{N+1} fb\cos\theta_i
    \label{eq:hamiltonian}.
\end{eqnarray}
The partition function of the system for one realization of quenched disorder (given by the sequence of occupation numbers) is
\begin{eqnarray}
    Z(n_1,\dots,n_N;N) &=& \prod_{i=1}^{N}\int\,d\Omega_i \exp\Biggr(\frac{fb \cos\theta_i}{k_BT}\Biggr)\nu(n_{i}) \nonumber \\
    &\times& \int\,d\Omega_{N+1} \exp\Biggr(\frac{fb \cos\theta_{N+1}}{k_BT}\Biggr)
    \label{eq:partit},
\end{eqnarray}
where we have defined
\begin{eqnarray}
    \nu(n_i)=\cases{1&for $n_i = 0$\\
    \frac{\delta(\theta_{i}-\theta_{i+1})\delta(\phi_{i}-\phi_{i+1})}{\sin\theta_i}&for $n_i=1.$} \\ \nonumber
    \label{eq:nu}
\end{eqnarray}

It is known that, in systems with quenched disorder, thermodynamic properties are calculated from  the disorder-averaged free energy (see, e.g,~\cite{fischer1993spin} and~\cite{dotsenko1995introduction}). Now, we consider the mean free energy averaged over all realizations of hinge disorder. Since the free energy for a specific realization is \(F=-\beta^{-1} \ln Z\) (\(\beta=1/k_BT\)), the mean free energy becomes
\begin{eqnarray}
    \langle F\rangle_N=-\beta^{-1}\sum_{n_1=0}^{1} \dots\sum_{n_N=0}^{1} (1-p)^{N-\sum n_i}p^{\sum n_i}\ln Z(n_1,\dots,n_N;N)
    \label{eq:exact_f}.
\end{eqnarray}
It can be trivially shown that if \(p=0\), it reduces to the free energy for the uFJC.

\section{\label{sec:exact}Exact Results} 

\subsection{\label{sec:toy_sys}Toy System}

Instead of heading to the thermodynamic limit immediately, it is helpful to consider a toy system consisting of a few hinges to get a grasp of the situation. However, because of the finite size of the system, the free energy and the extension of the toy system are not self-averaging. (The problem of self-averaging in finite systems is discussed later, in Section \ref{sec:self_avg}.) Thus, it may be risky to examine the behavior of such a system, but it is still useful to gain insight into the effect of quenched disorder. In this subsection, we consider a system with \(N=1\), which is sufficient for our purpose.

The mean free energy for the qFJC with \(N=1\) (free energy averaged of quenched disorder) reads
\begin{eqnarray}
    \langle F \rangle_1=-\Biggr[ p\ln\Biggr(\frac{4\pi \sinh(2f)}{2f} \Biggl)+(1-p)\ln \Biggr(\biggr(\frac{4\pi \sinh f}{f}\biggl)^2\Biggl)\Biggl]
    \label{eq:exact_f_1},
\end{eqnarray}
where we have set \(b=1\),\(\,\beta=1\) for simplicity. Notice that if \(p=0\), it reduces to the free energy of the uFJC as we have mentioned in~(\ref{eq:exact_f}). The force-extension relation in the Gibbs ensemble is given by
\begin{eqnarray}
    \frac{\langle z \rangle_1}{L}=-\frac{1}{L}\frac{\partial\langle F \rangle_1}{\partial f}=p \,\mathcal{L}(2f) +(1-p) \, \mathcal{L}(f)
    \label{eq:exact_ext_1},
\end{eqnarray}
where \(\mathcal{L}(x)=\coth x - x^{-1}\) is the Langevin function. Thus, like other problems with quenched disorder, the response to the force will be the average of the response. In our case, our goal is to compute the average over all possible combinations of the rods with various lengths, where the possible combinations of the lengths and the ways to partition integers considering the order have a one-to-one correspondence.

Moreover, since the response function becomes the average over combinations of the Langevin functions, it is obvious that the resulting response function for arbitrary \(N\) will still be a linear combination of Langevin-like functions, unlike the rFJC, also in the thermodynamic limit.~\cite{Noh_rFJC,Yi_rFJC} (In the case of rFJC, \(p\) increases when the force is increased. This results in a crossover to the state with longer rods, where the aligning becomes more favored.)

\subsection{\label{sec:recurrence}Recurrence Relation}

Even though we know the essence of the problem, the exact mean free energy, given in~(\ref{eq:exact_f}) cannot be cast into a closed form. Nevertheless, we can obtain a useful relation that allows the exact numerical calculation of the mean free energy for a finite number of hinges.

The mean free energy of \(N\) hinges can be expressed in terms of \(\langle F \rangle_k\), where \(k=0,\, 1\,, \dots,\, N-1\). First, consider the case when the first hinge, \(i=1\), is open. In that case the contribution to \(\langle F \rangle_N\) becomes
\begin{eqnarray}
    (1-p)\,p^0\Biggr[-\ln \Biggr( \frac{4\pi \sinh f}{f} \Biggl)+\langle F \rangle_{N-1} \Biggl].
\end{eqnarray}
Next, consider that the first hinge is closed and the second hinge is open. The corresponding contribution reads
\begin{eqnarray}
    (1-p)\,p^1\Biggr[-\ln \Biggr( \frac{4\pi \sinh 2f}{2f} \Biggl)+\langle F \rangle_{N-2} \Biggl].
\end{eqnarray}
Repeating this process to the \(N\,\)th hinge (the first \(N-1\,\)hinges are closed and the \(N\,\)th hinge is open) and collecting all the contributions, it can be written in a compact form:
\begin{eqnarray}
    (1-p)\sum_{k=0}^{N-1}p^k \, \Biggr[-\ln \Biggr(\frac{4\pi \sinh(k+1)f}{(k+1)f}\Biggl)+\langle F\rangle_{N-1-k}  \Biggl].
    \label{eq:recur_1}
\end{eqnarray}
Lastly, when all the hinges are closed, it trivially gives the contribution
\begin{eqnarray}
    -p^N\ln\Biggr(\frac{4\pi \sinh(N+1)f}{(N+1)f}\Biggl).
    \label{eq:recur_2}
\end{eqnarray}
Combining~(\ref{eq:recur_1}) and~(\ref{eq:recur_2}) leads to the recurrence relation of the mean free energy
\begin{eqnarray}
    \langle F \rangle_N=(1-p)\sum_{k=0}^{N-1}p^k \, \Biggr[&-&\ln \Biggr(\frac{4\pi \sinh(k+1)f}{(k+1)f}\Biggl)+\langle F\rangle_{N-1-k}  \Biggl]\nonumber\\&-&p^N\ln\Biggr(\frac{4\pi \sinh(N+1)f}{(N+1)f}\Biggl)
    \label{eq:recur},
\end{eqnarray}
where the initial condition is \(\langle F \rangle_0=-\ln(4\pi \sinh f/f)\) and \(N\geq1\).

\begin{figure}
    \centering
    \includegraphics[width=0.7\linewidth]{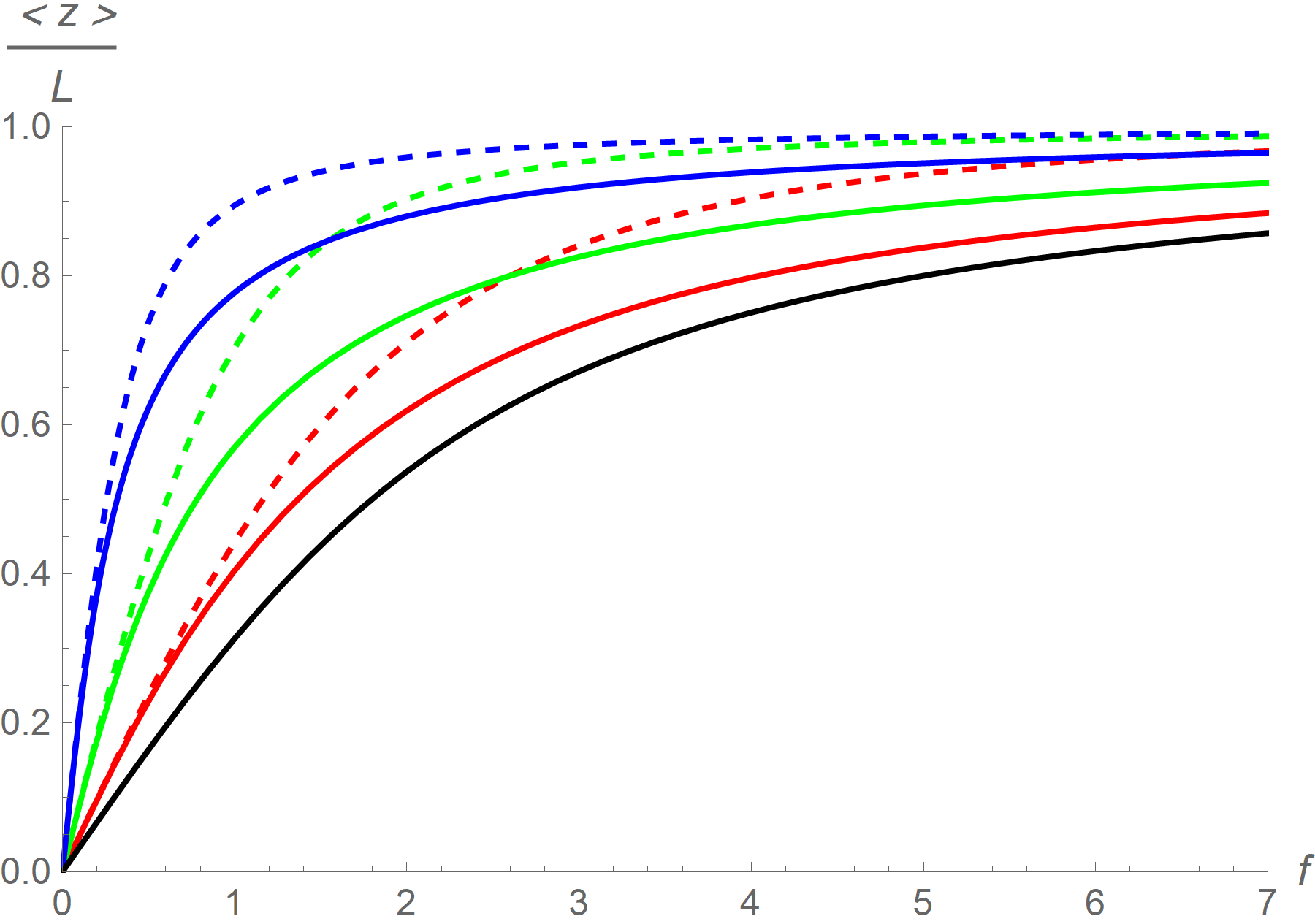}
    \caption{The extension of the qFJC, and the rFJC as a function of a force. The solid and the dashed curves refer to the qFJC and the rFJC. The red, green, and blue curves refer to \(p=0.2\), \(p=0.5\), and \(p=0.8\), respectively, and the black curve is for the uFJC, which is the Langevin function \(\mathcal{L}(f)\). \(N=17\), \(k_BT=1\), \(b=1\).}
    \label{fig:qFJC_ext}
\end{figure}

Using the recurrence relation obtained in~(\ref{eq:recur}), it is possible to calculate the exact stretching response for a finite number of hinges numerically. The extension of \(N=17\) qFJC is plotted in figure~\ref{fig:qFJC_ext}. As expected from Section~\ref{sec:toy_sys}, we observe that the qFJC is less extended than the rFJC under the same value of \(p\). Note that we have used \(p={\rm e}^\epsilon/({\rm e}^\epsilon+4 \pi)\) for the rFJC ($\epsilon$ being the activation energy for opening a reversible hinge), which is the closing probability when \(f=0\). Of course, both the rFJC and the qFJC have larger extensions than the uFJC for the same force, due to the presence of longer rods. 

\subsection{\label{sec:self_avg}Self-Averaging of Finite Systems}
Self-averaging plays an important role in describing a system with quenched disorder by its mean value. For random polymers, self-averaging of a certain property implies independence of that property on the random sequence (realization of disorder). In the thermodynamic limit, most of the physical quantities of a system without long-range correlations, such as the qFJC, are self-averaging due to the central limit theorem. However, for finite systems or near criticality, the problem of self-averaging becomes nontrivial. Self-averaging in random copolymers modeled as self-avoiding random walks has been studied by Whittington and collaborators. \cite{Orlandini_2002_Whittington,Soteros2004,james2002extent} They have derived a bound on the extent of self-averaging as a function of the polymer length. In this subsection, we estimate a minimal length of the qFJC beyond which self-averaging is a good approximation.

In principle, to be (at least approximately) self-averaging, the system should be large enough to realize most of the possible states that the system can have. Thus, it would be natural to try to find a minimal length that is sufficient to realize most of the possibilities. In our model, this length will be characterized by the closing probability of the hinge \(p\).

Let us denote this minimal length as \(L_m\) and define it as the contour length over which the polymer is likely to have at least two blocks, given that the probability of having a closed hinge is \(p\). This definition can be written as follows.
\begin{eqnarray}
    {\rm Prob}({\rm at \;least\;two\;blocks})>\frac{1}{2} \;\; {\rm or} \;\; {\rm Prob}({\rm one\;block})<\frac{1}{2}.
    \label{eq:selfavg_cond}
\end{eqnarray}
If a system has \(N\) hinges, the statement above can be written as \(p^N<1/2\), which is very simple. Thus, we conclude that the self-averaging is a good approximation if the number of hinges \(N\) satisfies
\begin{eqnarray}
    Nb>L_m=-\frac{\ln2}{\ln p}b.
\end{eqnarray}

We point out that this is just a rough estimate that illustrates the dependence of the minimal length on $p$. One may consider a different requirement for the number of blocks, for example, \mydoubleq{at least X blocks}. However, that would yield a transcendental equation.

\section{\label{sec:thm_lim}Thermodynamic Limit}

In this section, we consider \(N\rightarrow \infty\), which is the thermodynamic limit. To be more specific, we consider \(L\gg L_p\) (\(L_p\) is the persistence length) so that the orientational correlation vanishes at the end of the chain, {\it i.e.} no giant polymer block (cluster) exists. In this case, we expect that the bending becomes equivalent to stretching. A more detailed discussion about the equivalence of stretching and bending in the thermodynamic limit of rFJC is discussed in \cite{Yi_rFJC}.

\subsection{\label{sec:formalism}Formalism}

Consider a probability distribution of the polymer block (rod) to have \(k\) elementary segments, which we denote as \(\mathcal{P}(k)\), where \(k\) can be any value depending on the system. This probability distribution function \(\mathcal{P}(k)\) could also be viewed as the weighting function when the length \(kb\) polymer block is realized.

Now, consider we have \(N_B\) such blocks with no interaction between the blocks. Let the free energy of the length \(kb\) polymer block be given by \(\mathcal{F}(k)\). This yields the mean free energy of the whole chain as
\begin{eqnarray}
    \langle F_{\rm{chain}}\rangle = N_B \sum_{k} \mathcal{P}(k) \mathcal{F}(k)
    \label{eq:mean_free_chain},
\end{eqnarray}
where the mean operation is done over the realizations of the ensemble. If we consider \(N_B \gg 1\), implying that there is no giant block, the number of monomers of the whole chain will be \(N = S_BN_B\), where \(S_B=\sum_{k} k \,\mathcal{P}(k)\) is the mean block size averaged over blocks, {\it not the monomers}. This leads to the mean free energy density per monomer
\begin{eqnarray}
    \langle F_{\rm density} \rangle = \frac{\langle F_{\rm chain}\rangle}{N} = \frac{1}{S_B} \sum_{k} \mathcal{P}(k) \mathcal{F}(k)
    \label{eq:mean_free_density}.
\end{eqnarray}

In general, one can obtain the exact mean free energy density from~(\ref{eq:mean_free_density}) for any system if there is no interaction between blocks. However, the real problem is that it is hard to compute the sum in most cases. In the next sections, we will show how it can be used for the qFJC in the thermodynamic limit.

\subsection{\label{sec:discrete}Discrete qFJC}

To make~(\ref{eq:mean_free_density}) useful, knowing both \(\mathcal{P}(k)\) and \(\mathcal{F}(k)\) is necessary. For the qFJC, \(\mathcal{F}(k)\) is simple, and it is given as
\begin{eqnarray}
    \mathcal{F}(k) = -\beta^{-1} \ln \Biggl( \frac{4 \pi \sinh(\beta kbf)}{\beta kbf} \Biggr)
    \label{eq:block_fenergy}.
\end{eqnarray}
Let us now focus on \(\mathcal{P}(k)\). To form a length a block of length \(kb\), we should have \(k-1\) hinges consecutively closed, and the \(k\,\)th hinge should be open. Assuming that the opening-closing of each hinge as independent events, it leads to the weighting function
\begin{eqnarray}
    \mathcal{P}(k) = p^{k-1} \, (1-p)
    \label{eq:disqFJC_block_weight}.
\end{eqnarray}
This determines the value \(S_B=1/(1-p)\), the mean block size averaged over blocks.

Before we proceed, note that the weighting function \(\mathcal{P}(k)\) has a similar form (except for a multiplicative constant) to the orientational correlation function for the discrete chain, which is calculated in \citeonline{Yi_rFJC}. This is somehow expected. If we define $\vec{u}(i)$ as the tangent vector of the \(i\)th segment, $\vec{u}(i)\cdot\vec{u}(j)$ can have two values: $0$ or $1$. For the latter, the hinges between the \(i\)th segment and the \(j\)th segment should be consecutively closed, which corresponds to the case where the size \(|i-j|+1\) block has formed. The former corresponds to the other cases, which give zero contribution to the orientational correlation function. Thus, we have \(\langle\vec{u}(i)\cdot\vec{u}(j)\rangle \sim \mathcal{P}(|i-j|)\).

Now we are ready to use~(\ref{eq:mean_free_density}). Substituting~(\ref{eq:block_fenergy}) and (\ref{eq:disqFJC_block_weight}), we end up with
\begin{eqnarray}
    \langle F_{\rm{density}} \rangle = -\sum_{k=1}^{\infty} (1-p)^2 \, p^{k-1} \ln \Biggl( \frac{4 \pi \sinh(kf)}{kf} \Biggr)
    \label{eq:disqFJC_fenergy},
\end{eqnarray}
where we have used \(\beta=1\), \(b=1\). However, one can easily notice that it is impossible to compute the sum exactly. Instead, using the fact that the force-extension relation is given as \(\langle z \rangle = -\partial_f \langle F \rangle\) in the Gibbs ensemble, the force-extension relation is
\begin{eqnarray}
    \frac{\langle z \rangle}{L} = \sum_{k=1}^{\infty} (1-p)^2 \, p^{k-1} k\, \mathcal{L}(kf)
    \label{eq:disqFJC_ext}.
\end{eqnarray}
Notice that the resulting \mydoubleq{normalized} weighting function \((1-p)^2 p^{k-1}k\) is the probability of a randomly chosen site being a part of a size \(k\) cluster in the 1-D percolation theory.~\cite{aharony2003percolation}
Still, it is clear that it cannot be computed exactly, but an approximation for some limits will be useful.

First, in the small force limit, the series expansion yields
\begin{eqnarray}
    \frac{\langle z \rangle}{L} &=& \frac{(1-p)^2}{p}\sum_{n=1}^{\infty} \frac{1}{f}\frac{B_{2n}(2f)^{2n}}{(2n)!} \,{\rm{Li}}_{-2n}(p)  \nonumber \\
    &=& \frac{(1+p)\,f}{(1-p)\,3} + \mathcal{O}(f^3) 
    \label{eq:disq_small},
\end{eqnarray}
where \({\rm{Li}}_{s}(z)\) is the polylogarithm of order \(s\) and \(B_n\) is the Bernoulli number. Notice that \((1+p)/(1-p)\) is the mean cluster size averaged over sites in one dimension, and we have the same linear response as in the rFJC since the force does not affect the opening-closing of hinges when \(f=0\). This can also be observed numerically in figure~\ref{fig:qFJC_ext}.

Now consider the large force limit, where \(\mathcal{L}(x)\approx 1-1/x\). Substituting this expression to~(\ref{eq:disqFJC_ext}) leads to
\begin{eqnarray}
    \frac{\langle z \rangle}{L} &\approx& 1-\frac{1-p}{f}
    \label{eq:disq_large}.
\end{eqnarray}
By analogy to the force-extension relation of the uFJC at the strong-force limit, we see that the response expressed by (\ref{eq:disq_large}) is similar to that of a uFJC with monomer (elementary rod) length equal to $b/(1-p)$. Notice that the response in the two regimes ((\ref{eq:disq_small}) and~(\ref{eq:disq_large})) can be expressed by that of a uFJC with two different sizes of effective monomers, \((1+p)b/(1-p)\) and \(b/(1-p)\), respectively.

We see that the resulting force-extension relation qualitatively behaves similar to a Langevin function, in the sense that the linear response is proportional to the mean cluster size. However, it approaches full extension with an inverse-force decay with an effective monomer size which is different from the mean cluster size. Thus, it is not exactly true that the tensile elasticity of a qFJC reduces to that of a uFJC with an effective monomer size. 

We can prove that the extension of the rFJC is larger than in the qFJC by arguing that both force-extension relations are monotonic functions of the force. Since the extension of the rFJC approaches 1 with an exponential decay and both the qFJC and the rFJC have the same linear response, that automatically implies that the extension is larger than the qFJC under the same conditions.

\subsection{\label{sec:cont}Continuous qFJC}

\begin{figure}
    \centering
    \includegraphics[width=0.6\linewidth]{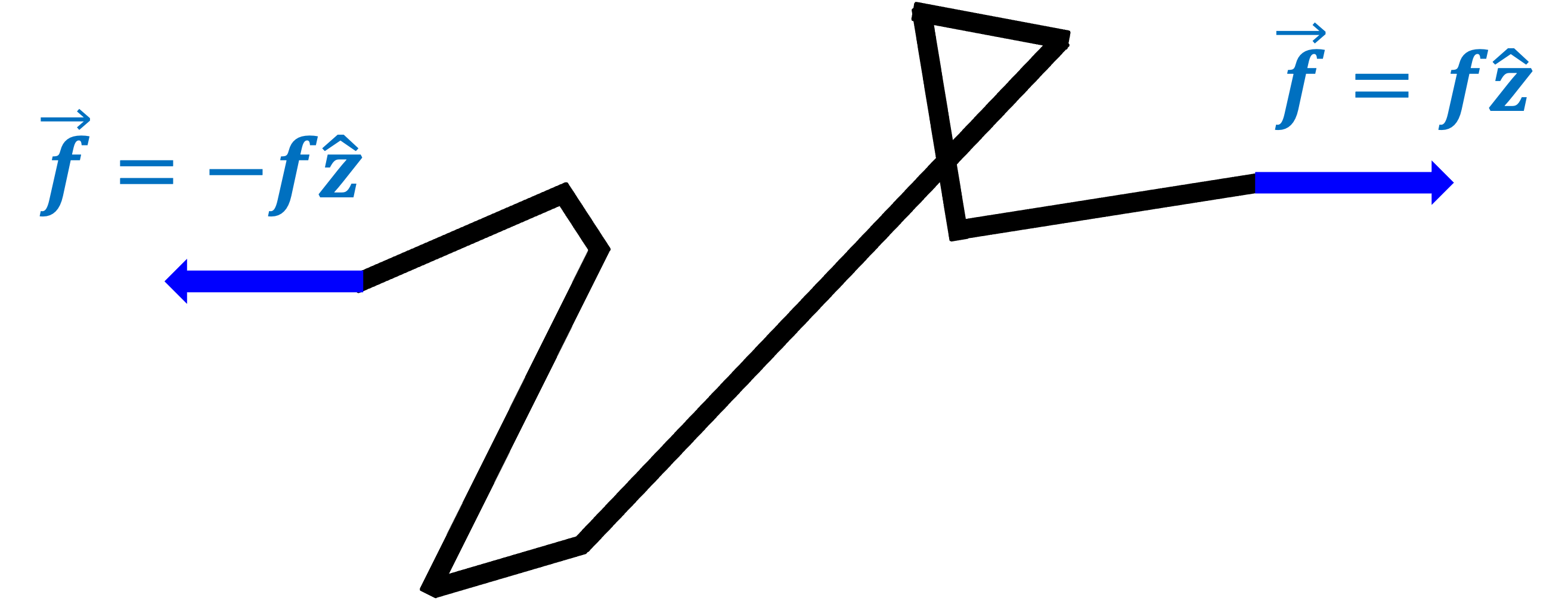}
    \caption{One realization of the continuous qFJC under tensile force.}
    \label{fig:cont_qFJC}
\end{figure}

One generalization of the quenched Freely Jointed Chain is considering the continuum limit, which is also discussed in~\cite{Yi_rFJC}. A specific realization of such a chain is drawn in figure~\ref{fig:cont_qFJC}. For the continuum limit, we have the weighting function as \(\mathcal{P}(l)={\rm e}^{-l/L_p}/L_p\), where \(l\) is the block length and \(L_p\) is the persistence length. As we discussed in Section \ref{sec:discrete}, this weighting function can be justified by considering the orientational correlation function for the continuum limit, which has the form of \(\exp(-L/L_p)\).

The free energy of a length \(l\) block, \(\mathcal{F}(l)\) is given by~(\ref{eq:block_fenergy}), with changing the variable as \(l=kb\). Then, the force-extension relation reads
\begin{eqnarray}
    \frac{\langle z \rangle}{L} = \frac{1}{{L_p}^2}\int_{0}^{\infty} {\rm e}^{-l/L_p} \mathcal{L}(lf)\, l \,dl
    \label{eq:cont_ext_integral}.
\end{eqnarray}

\begin{figure}
    \centering
    \includegraphics[width=0.7\linewidth]{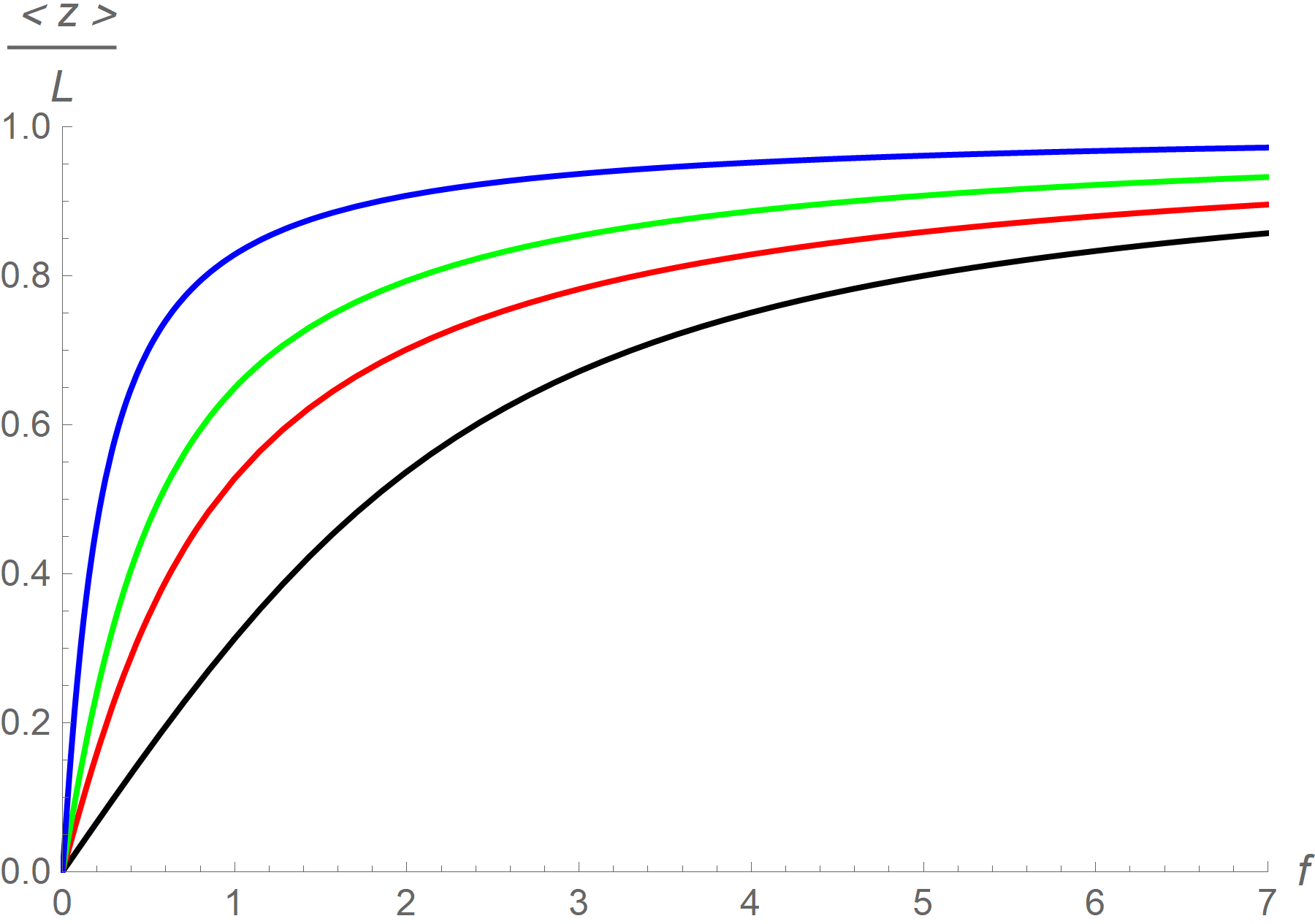}
    \caption{The extension of the continuous qFJC as a function of a force. The red, green, and blue curves refer to \(L_p=1/(1-0.2)=1.25\), \(L_p=1/(1-0.5)=2\), \(L_p=1/(1-0.8)=5\). The black curve is the extension of the uFJC. \(k_BT=1\), \(b=1\).}
    \label{fig:cont_qFJC_ext}
\end{figure}

Surprisingly, this non-trivial integral can be cast into a relatively simple form by some manipulations, and using a result of~\cite{zwillinger2007table}, we obtain
\begin{eqnarray}
    \frac{\langle z \rangle}{L} = \alpha^2 \Biggl[ \frac{1}{2}\psi'\Biggl( \frac{\alpha}{2} \Biggr) - \frac{1}{\alpha} - \frac{1}{\alpha^2} \Biggr]
    \label{eq:cont_ext_exact},
\end{eqnarray}
where \(\alpha=(L_pf)^{-1}\) and \(\psi'(z)=\partial^2_z \ln \Gamma(z)\), the trigamma function. $\Gamma(z)$ is the gamma function. The plot of~(\ref{eq:cont_ext_exact}) is shown in Figure~\ref{fig:cont_qFJC_ext}.

It is also interesting to examine the behavior at the limits of small and large forces. For the former, where \(\alpha/2 \rightarrow \infty\), by using the expansion of the gamma function, we obtain
\begin{eqnarray}
    \frac{\langle z \rangle}{L} &=& \sum_{n=1}^{\infty} \frac{1}{f}\frac{B_{2n}(2f)^{2n}}{(2n)!} \,\frac{(2n)!}{(1-p)^{2n-1}} \nonumber \\
    &=& 2L_p\frac{f}{3}-24{L_p}^3\frac{f^3}{45}+\mathcal{O}(f^5)
    \label{eq:cont_small}.
\end{eqnarray}
Note that \(L_p=b/(1-p)\) in the continuum limit.~\cite{Yi_rFJC} The result above can be confirmed by the linear response theory, which states that \(\langle z \rangle \propto \langle z^2 \rangle f\) where \(\langle z^2 \rangle \approx 2L_pL/3 \; (L \gg L_p)\) is the mean squared end-to-end distance.~\cite{Yi_rFJC} Note also that
\begin{eqnarray}
    \frac{(1-p)^2}{p} {\rm Li}_{-2n}(p) \approx \frac{(2n)!}{(1-p)^{2n-1}}
    \label{eq:polylog_approx},
\end{eqnarray}
when \(p \rightarrow 1\).~\cite{arfken2011mathematical} This implies that the result of the continuum limit becomes identical to that of the discrete chain at the limit of \(p \rightarrow 1\). This is somehow expected from the fact that the continuum limit implies \(N \rightarrow \infty\) and \(b \rightarrow 0\) but keeps \(L_p=b/(1-p)\) constant.

In the large-force regime (\(|\alpha/2| < 1\)), we have
\begin{eqnarray}
    \frac{\langle z \rangle}{L} = 1-\frac{1}{L_p f}+\frac{\zeta(2)}{2{L_p}^2f^2}+\mathcal{O}\Biggl(\frac{1}{f^3}\Biggr)
    \label{eq:cont_large},
\end{eqnarray}
where \(\zeta(z)\) is the Riemann zeta function. Note that, also in this case, it cannot be viewed as an extension of a uFJC with an effective monomer length of size \(L_p\), due to the continuous nature of this model. Moreover, even if we try to express it as an uFJC, those two regimes have different effective monomer lengths, \(2L_p\) (the Kuhn length of a flexible chain) and \(L_p\).

\section{\label{sec:general}Generalization}

\subsection{\label{sec:gen_form}Copolymers with Non-Interacting Blocks}

The expression for the free energy density in the thermodynamic limit~(\ref{eq:mean_free_density}) is for homopolymers with non-interacting blocks. The quenched disorder is related to the size of the blocks that follow a certain random distribution. We can generalize our analysis by considering two or more types on non-interacting random blocks, each following a different distribution.
%Moreover, if we assume that the interaction between the blocks is sufficiently weak, it can be generalized to random copolymers.

Consider a sequence of blocks, {\it{e.g.}}, 1-2-1-2- ... -1-2, with different block lengths. It can be viewed as a block copolymer comprising two types of blocks, type 1 and type 2, with \(N_B\) blocks each. If the length distribution of the type \(i\) block follows a probability distribution function \(\mathcal{P}_i(k)\), the mean free energy of the chain reads
\begin{eqnarray}
    \langle F_{\rm{chain}}\rangle = N_B  \sum_{k} \bigl( \mathcal{P}_1(k) \mathcal{F}_1(k) + \mathcal{P}_2(k) \mathcal{F}_2(k) \bigr)
    \label{eq:mean_free_chain_copolymer}.
\end{eqnarray}
In this case, the number of monomers in the chain will be \(N=(S_1+S_2)N_B\), where \(S_1\) and \(S_2\) are the mean sizes of blocks of type 1 and type 2, respectively. This leads to the mean free energy density
\begin{eqnarray}
    \langle F_{\rm{density}} \rangle = \frac{ \sum_{k}\mathcal{P}_1(k) \mathcal{F}_1(k) + \sum_{k} \mathcal{P}_2(k) \mathcal{F}_2(k)}{S_1+S_2}
    \label{eq:mean_free_density_copolymer}.
\end{eqnarray}
If we have \(J\) types of polymer blocks, the mean free energy density can be trivially written as
\begin{eqnarray}
    \langle F_{\rm{density}} \rangle = \frac{\sum_{j=1}^J  \sum_{k} \mathcal{P}_j(k) \mathcal{F}_j(k)}{\sum_{j=1}^J S_j}
    \label{eq:mean_free_density_copolymer_generalized}.
\end{eqnarray}

Now we consider the force-extension relation of the polymer. In the Gibbs ensemble, the extension reads
\begin{eqnarray}
    \frac{\langle z \rangle}{L} &=& -\,\frac{\sum_{j=1}^J  \sum_{k} \mathcal{P}_j(k) \, \partial_f \mathcal{F}_j(k)}{S} \nonumber \\
    &=& \sum_{j=1}^J  \,\frac{S_j  \langle z \rangle_j}{S L} \label{eq:mean_ext_generalized} \\
    &=& \sum_{j=1}^J\, \theta_j \frac{\langle z \rangle_j}{L} \nonumber,
\end{eqnarray}
where we have used \(S=\sum_j S_j\), \(\theta_j=S_j/S\), and \(\langle z \rangle_j\) as the extension of homopolymer of type \(j\). Notice that \(\theta_j\) is the fraction of monomers of type \(j\) in the whole chain. (\ref{eq:mean_ext_generalized}) implies that the resulting response of the random copolymer to the tension is the mean value of the corresponding response of each block type, which also agrees with our intuition. 
%Note also that a similar argument can be made in the Helmholtz ensemble.

\subsection{\label{sec:hc-trans}Example: Rod-Coil Multiblock Copolymers}

\begin{figure}
    \centering
    \includegraphics[width=0.7\linewidth]{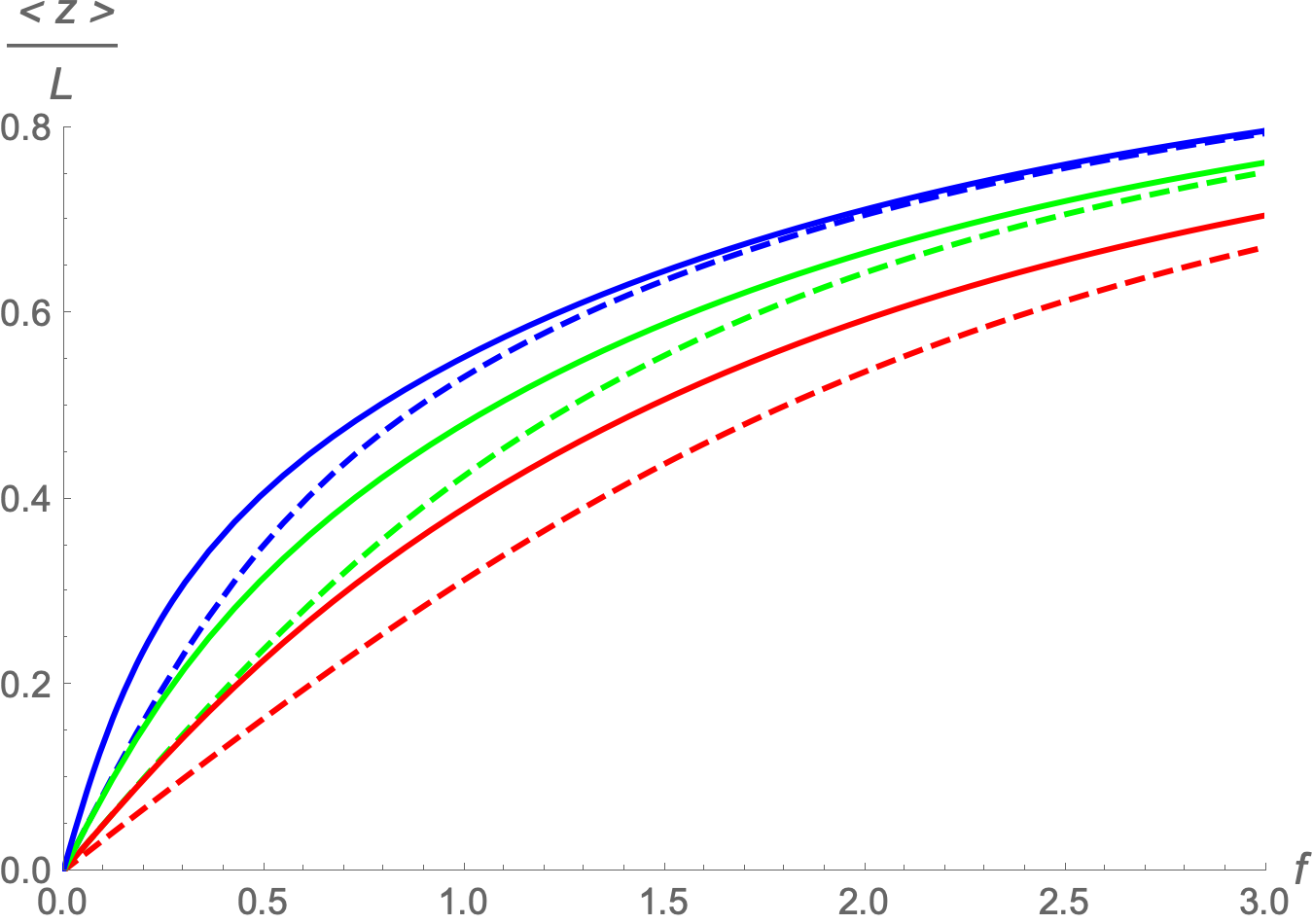}
    \caption{The extension of the rod-coil multiblock copolymer as a function of a force. The solid and the dashed curves refer to the exact value and the monodisperse-rod (mean-field) approximation, respectively. The red, green, and blue curves refer to \(L_p=1\), \(L_p=2\), and \(L_p=4\), respectively. \(\theta=0.5\), \(k_BT=1\), \(b=1\)}
    \label{fig:rod_coil_ext}
\end{figure}

An interesting example of the implementation of this generalization is the rod-coil multiblock copolymers. As in~\cite{Buhot_PRL}, consider that a copolymer consists of two types of blocks: the rod and the coil. Let the length distribution of the rod and the coil be given by \(\mathcal{P}_{\rm{rod}}(k)\) and \(\mathcal{P}_{\rm{coil}}(k)\) respectively. The free energy of a rod and a coil of size (number of monomers) $k$ is
\begin{eqnarray}
    \mathcal{F}_{\rm{rod}}(k) &=& \,-\ln \Biggl( \frac{4 \pi \sinh(kf)}{kf} \Biggr) \nonumber \\
    \mathcal{F}_{\rm{coil}}(k) &=& -k \ln \Biggl( \frac{4 \pi \sinh(f)}{f} \Biggr)
    \label{eq:copolymer_block_fenergy},
\end{eqnarray}
respectively.
Note that we have used the free energy of a rod of $k$ monomers for \(\mathcal{F}_{\rm{rod}}(k)\) and the free energy of the usual FJC (uFJC) for \(\mathcal{F}_{\rm{coil}}(k)\). This yields the extension as
\begin{eqnarray}
    \frac{\langle z \rangle}{L} &=& \frac{1}{S_{\rm rod }+S_{\rm coil}} \Biggl[ \, \sum_{k} \mathcal{P}_{\rm{rod}}(k)k\,\mathcal{L}(kf) + \sum_{k}\mathcal{P}_{\rm{coil}}(k)k\,\mathcal{L}(f) \Biggr] \nonumber \\ \\
    &=& \frac{\sum_{k} \mathcal{P}_{\rm{rod}}(k)k\,\mathcal{L}(kf)}{S_{\rm{rod}}+S_{\rm{coil}}} + \frac{S_{\rm{coil}}\,\mathcal{L}(f)}{S_{\rm{rod}}+S_{\rm{coil}}} \nonumber
    \label{eq:copolymer_block_ext}
\end{eqnarray}

If we consider a monodisperse distribution of the rod size such that \(\mathcal{P}_{\rm{rod}}(k)=\delta(k-S_{\rm{rod}})\), it yields the same result as in~\cite{Buhot_PRL}. (One could view this monodisperse distribution as a kind of mean-field approximation, where we consider all rods to have the same size and neglect fluctuations.) Defining the fraction of monomers to be in the helix (rod) state as \(\theta = S_{\rm{rod}} /(S_{\rm{rod}}+S_{\rm{coil}})\), the force-extension relation can be written by
\begin{eqnarray}
    \frac{\langle z \rangle}{L} = \theta \, \mathcal{L}(S_{\rm{rod}} \,f) +(1-\theta)\mathcal{L}(f)
    \label{eq:copolymer_block_ext_PRL}.
\end{eqnarray}
The difference in the notation with~\cite{Buhot_PRL} comes from the fact that we have considered that the rod consists of \(S_{\rm{rod}}\) monomers, whereas the reference has considered a single large monomer of size \(S_{\rm{rod}}\).

However, the form of \(\mathcal{P}_{\rm{rod}}(k)\) can be different from the monodisperse and that yields a different force-extension relation, even though the value of \(\theta\) remains the same. One example is a rod treated as a continuous polymer with an exponential length distribution, an element of the continuous qFJC that we discussed above. The comparison between the exact result and the monodisperse (mean-field) approximation for \(\theta=0.5\) is drawn in figure~\ref{fig:rod_coil_ext}. It turns out that the two different distributions of the block size yield different force-extension relations. Since the longer rods align with the direction of the force more easily, the mean extension increases in the small and also in the intermediate-force regimes. In the large-force regime, the discrepancies become negligible since most of the segments are already aligned to the direction of the force.

\section{\label{sec:conclusion}Conclusion and Discussion}

In this article, we have introduced and analyzed the elastic behavior of the qFJC, a freely-jointed chain with quenched disorder in the closed-open state of its hinges. 
The qFJC captures the quenched disorder in the local bending stiffness of the polymer. We calculated the response by averaging the response over all realizations of the disorder, as in other systems with quenched disorder. Note that there is a one-to-one correspondence between the possible configurations of the qFJC and the ordered partitioning of integers. This can be understood if we consider the hinges as the possible sites of the partitions. Then, the number of segments between the partitions becomes the size of the integer.

The recurrence relation that we derived for the mean free energy allows us to calculate the exact force-extension relation numerically, although it becomes inefficient for a large value of random hinges \(N\). We expect our results to be approximately self-averaging, if the contour length is longer than a certain minimal length \(L_m\) that depends on the closing probability of a hinge $p$. As we expected, it turns out that the force-extension relation is still qualitatively similar to a Langevin function, with a linear response at small forces and an inverse-force limit at strong stretching.
%no non-monotonic behavior for the compliance and the extension. 
Compared to the rFJC, the qFJC is always less extended than the rFJC for any value of the stretching force \(f\).

We have derived the exact mean free energy density in the thermodynamic limit, making an analogy with the percolation theory. The result is expressed as an infinite series, and in most cases, it turns out that it is hard to use it directly. However, we obtain simple expressions at the linear-response and the strong-stretching limits. The former is the same as in the rFJC. The latter is qualitatively different from the rFJC and behaves as a uFJC with an effective monomer length. We should point out that both (\ref{eq:disqFJC_ext}) and (\ref{eq:cont_ext_integral}) could be "derived" directly, without any detour through free energy calculations, by considering a weighted superposition of Langevin functions.

The mathematical form we have derived for the discrete qFJC is general and can also be applied to the continuum limit. Remarkably, the continuum limit of the qFJC yields a closed expression for the force-extension relation, involving the trigamma function. 
%Since the force-extension relation becomes the function of the digamma function, it allows us to calculate the exact extension for arbitrary \(f\) and \(p\) numerically in the thermodynamic limit.

As expected, the projection of the mean end-to-end distance on the axis of the tensile force for the qFJC is always less extended than that of the rFJC for both the discrete and the continuous case, also in the thermodynamic limit. In the reversible chain (rFJC), there is a crossover to the state with larger bending stiffness (all reversible hinges getting closed) when the force increases, resulting in the formation of a longer rod with a larger response.~\cite{Yi_rFJC,Noh_rFJC} In contrast, since the disorder architecture is quenched, we do not have such a crossover. This is crucial for understanding the smaller response for the qFJC.

If we want to experimentally determine the type of disorder for this kind of system, we must focus on the large-force regime. Although the initial linear response of the qFJC and the rFJC is the same, the two models behave qualitatively differently in the large-force regime. For both discrete and continuous qFJC, the chain extension approaches 1 (full extension) as \(1-A/f\), whereas the rFJC approaches 1 as \(1-B\exp(-kf)\) (where \(A,\,B\), and \(k\) are constants) in the thermodynamic limit.

Noticing that the derivation of~(\ref{eq:mean_free_density}) is based on the assumption of non-interacting blocks, we generalized to copolymers consisting of different types of random blocks, each having a different length distribution. We have shown that the resulting force-extension relation becomes the average of the extension function over all types of random blocks. When we extended our analysis to model rod-coil multiblock copolymers, it turned out that the length distribution of the rod affects the force-extension relation, especially in the small-force regime, compared to the case of monodisperse rods.

The rFJC and qFJC can be viewed as minimal models of polymers where the local bending stiffness fluctuates along the polymer backbone. For example, some DNA-binding proteins may bind/unbind along the single- or double-stranded DNA and increase its local bending stiffness. \cite{DNA_Protein_Bruinsma, DNA_Protein_experiments, Andelman_PRE} If the typical time scale of the binding/unbinding process is smaller than the relaxation time of the DNA, the rFJC will be a more appropriate model. In the opposite case, the qFJC may be more useful. In DNA(or mRNA) that undergoes translation(transcription), the time scale of the translation(transcription) is separated from the relaxation time of the polymer. \cite{ALHASHIMI2008321} The presence of a ribosome(polymerase) on the polymer backbone increases its local bending stiffness. For a time scale smaller than the typical transport time of such a process, the qFJC may be a useful model to describe the tensile elasticity.  

\section*{Acknowledgments}
We thank Juhyeon Lee for his useful comments and discussion. We acknowledge the support of a grant from the National Research Foundation of Korea, NRF-2022R1F1A1070341, funded by the Ministry of Science and ICT, Korea (MSIT).

\section*{References}

\bibliography{qFJC}

\end{document}